# United Monoids

## Finding Simplicial Sets and Labelled Algebraic Graphs in Trees

Andrey Mokhov[a,b]

a   Jane Street, London, UK

b   Newcastle University, Newcastle upon Tyne, UK

**Abstract**     Graphs and various graph-like combinatorial structures, such as preorders and hypergraphs, are ubiquitous in programming. This paper focuses on representing graphs in a purely functional programming language like Haskell. There are several existing approaches; one of the most recently developed ones is the "algebraic graphs" approach (2017). It uses an algebraic data type to represent graphs and has attracted users, including from industry, due to its emphasis on equational reasoning and making a common class of bugs impossible by eliminating internal invariants.

    The previous formulation of algebraic graphs did not support edge labels, which was a serious practical limitation. In this paper, we redesign the main algebraic data type and remove this limitation. We follow a fairly standard approach of parameterising a data structure with a semiring of edge labels. The new formulation is both more general and simpler: the two operations for composing graphs used in the previous work can now be obtained from a single operation by fixing the semiring parameter to zero and one, respectively.

    By instantiating the new data type with different semirings, and working out laws for interpreting the resulting expression trees, we discover an unusual algebraic structure, which we call "united monoids", that is, a pair of monoids whose unit elements coincide. We believe that it is worth studying united monoids in their full generality, going beyond the graphs which prompted their discovery. To that end, we characterise united monoids with a minimal set of axioms, prove a few basic theorems, and discuss several notable examples.

    We validate the presented approach by implementing it in the open-source `algebraic-graphs` library. Our theoretical contributions are supported by proofs that are included in the paper and have also been machine-checked in Agda. By extending algebraic graphs with support for edge labels, we make them suitable for a much larger class of possible applications. By studying united monoids, we provide a theoretical foundation for further research in this area.

## The Art, Science, and Engineering of Programming

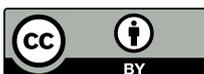

**United Monoids**

## 1 Introduction

Graphs and various graph-like combinatorial structures are ubiquitous in programming. There are numerous approaches to representing graphs, from good old adjacency matrices [4] to categorical graph algebras [34]. This paper introduces a new purely functional graph representation, building on the earlier work on *algebraic graphs* [21]. The new representation is based on elementary foundations (trees and semirings), and provides an expressive language for modelling graph-like structures.

The following data type of binary trees with a-labelled leaves and s-labelled internal nodes will be the main protagonist of our story[1]:

```
data Tree s a = Leaf a | Node s (Tree s a) (Tree s a)
```

In this paper we are going to interpret such trees in a few different ways. To do that, we will *fold* them [10] by specifying different pairs of functions leaf and node, which will give meanings (of type m) to the leaves and internal nodes, respectively:

```
fold :: (a -> m) -> (s -> m -> m -> m) -> Tree s a -> m
fold leaf node = meaning
  where
    meaning (Leaf a) = leaf a
    meaning (Node s x y) = node s (meaning x) (meaning y)
```

For example, to compute the *size* of a tree, i.e., the number of leaves in it, we can fold the tree into an Int as follows.

```
size :: Tree s a -> Int
size = fold (\a -> 1) (\s x y -> x + y)
```

Here we ignore the labels of leaves and internal nodes: every leaf is interpreted as 1, and every internal node simply adds up the sizes of its children. In the rest of the paper, we will study more interesting ways to fold trees. In particular, we will see how the very same data type can serve us when working with sets, preorders, various flavours of graphs, and simplicial sets. More specifically, our contributions are as follows.

- We extend *algebraic graphs* [21] to support edge labels (§7), and also simplify their inductive definition from four to just two constructors (the Tree data type, §2).
- To demonstrate the flexibility of the new approach, we show how it can be used to work with a variety of combinatorial structures, including simplicial sets and preorders (§3-§6). All of these seemingly different structures turn out to be just different interpretations of the same underlying language of trees.
- We introduce *united monoids* as an algebraic structure that captures the essence of the various studied interpretations of trees (§8). We also characterise united monoids with a minimal set of axioms and prove a few basic theorems.

We review related work in §9.

---

[1] We will use Haskell throughout this paper but the presented ideas are not Haskell-specific and can be readily translated to other languages.

## 2 Trees

In this section we study the Tree s a data type in more detail. An experienced functional programmer will find the data type and various associated functions fairly standard, so this section can be skipped after glancing through the definitions in Listings 1 and 2.

Throughout the paper, leaves of our trees will represent *elements* (of sets and preorders) or *vertices* (of simplicial sets and graphs). We will treat their labels a as abstract, apart from the occasional requirement of Eq a or Ord a, e.g., to be able to collect leaves into a set. Internal nodes, on the other hand, will represent various kinds of *connectivity*, and we will require that their labels s come from a semiring [12].

A *semiring* is an algebraic structure that generalises arithmetic: it's a set equipped with associative[2] operations of *addition* ⊕ and *multiplication* ⊗, whose units[3] are *zero* ⓪ and *one* ①, respectively. Semirings give us a basic language for connectivity:

- The lack of connectivity is denoted by ⓪. For instance, zero network bandwidth.
- Addition ⊕ is a commutative[4] operation for combining connections in *parallel*; e.g., selecting a link with the *maximum* bandwidth from available alternatives.
- Multiplication ⊗ combines connections in *sequence*. Continuing the example, to compute the bandwidth of a route, we take the *minimum* over all links that need to be traversed. Multiplication distributes[5] over addition and has ⓪ as its zero.[6]
- Finally, ① denotes either a unit of connectivity or infinite connectivity, depending on a specific semiring. In the bandwidth semiring, we have $① = \infty$.

We first turn our attention to trees whose internal nodes are labelled with ⓪, i.e., whose leaves are *disconnected*. For example, the trees shown below correspond to expressions "a", "b" ⋄ "b", ("a" ⋄ "b") ⋄ "c" and "a" ⋄ ("b" ⋄ "c"), where the operator ⋄ comes from the Semigroup (Tree s a) instance, and string literals are desugared into leaves using the OverloadedStrings extension (see Listing 1).

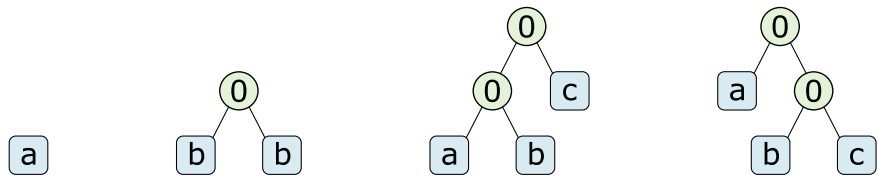

A *semigroup* is a set with an associative operation. It may seem worrying that our Semigroup instance yields different trees for left- and right-associated expressions, as in the last two examples. To resolve this, we make the Tree data type abstract and require all uses of fold to respect the associativity when interpreting trees. Concretely, the fold's argument node must yield associative functions for all s, including s = zero.

---

[2] An operation • is *associative* if $(a • b) • c = a • (b • c)$. Associativity is convenient both for humans (making parentheses unnecessary) and machines (allowing large expressions to be efficiently processed by partitioning them into smaller subexpressions as necessary).

[3] An element $e$ is the *unit* (or the *identity element*) of an operation • if $e • a = a • e = a$.

[4] An operation • is *commutative* if $a • b = b • a$.

[5] *Distributivity* of ⊗ over ⊕ means $a ⊗ (b ⊕ c) = (a ⊗ b) ⊕ (a ⊗ c)$ and $(a ⊕ b) ⊗ c = (a ⊗ c) ⊕ (b ⊗ c)$.

[6] An element $z$ is the *zero* (or the *annihilating element*) of an operation • if $z • a = a • z = z$.

■ **Listing 1** The Tree data type and instances of various standard Haskell type classes.

```haskell
data Tree s a = Leaf a | Node s (Tree s a) (Tree s a)

fold :: (a -> m) -> (s -> m -> m -> m) -> Tree s a -> m  -- Note: node s must be associative
fold leaf node = meaning
  where
    meaning (Leaf a) = leaf a
    meaning (Node s x y) = node s (meaning x) (meaning y)

class Semiring s where  -- In this paper, s in Tree s a is always a Semiring
    zero :: s           -- Unit of ⊕, zero of ⊗
    (⊕)  :: s -> s -> s -- Associative, commutative
    one  :: s           -- Unit of ⊗
    (⊗)  :: s -> s -> s -- Associative, distributes over ⊕

instance Semiring s => Semigroup (Tree s a) where
    (◇) :: Tree s a -> Tree s a -> Tree s a
    (◇) = Node zero  -- All tree interpretations must respect the associativity of ◇

instance Functor (Tree s) where
    fmap :: (a -> b) -> Tree s a -> Tree s b  -- Apply a function a -> b to every leaf a
    fmap f = fold (Leaf . f) Node

instance Applicative (Tree s) where
    pure :: a -> Tree s a  -- Create a "trivial" tree containing a single leaf
    pure = Leaf
    (<*>) :: Tree s (a -> b) -> Tree s a -> Tree s b  -- Graft the 2nd tree on every leaf of the 1st
    (<*>) = ap                                         -- Standard implementation via Monad

instance Monad (Tree s) where
    (>>=) :: Tree s a -> (a -> Tree s b) -> Tree s b  -- Graft a tree f a on every leaf a
    x >>= f = fold f Node x

instance Foldable (Tree s) where
    foldr :: (a -> b -> b) -> b -> Tree s a -> b  -- Fold leaves from right to left
    foldr f b tree = fold f (const (.)) tree b

instance Traversable (Tree s) where
    traverse :: Applicative f => (a -> f b) -> Tree s a -> f (Tree s b)  -- Effectful variant of fmap
    traverse f = fold (fmap Leaf . f) (liftA2 . Node)

instance IsString a => IsString (Tree s a) where  -- This makes "a" a shortcut for Leaf "a"
    fromString = Leaf . fromString
```

▪ **Listing 2**   A basic API for constructing and manipulating trees. We use the standard module Data.List.NonEmpty for representing non-empty lists, and Data.Set for sets.

```haskell
leaf :: a -> Tree s a -- A tree with a single leaf
leaf = Leaf

leaves :: Semiring s => NonEmpty a -> Tree s a -- A tree of "disconnected" leaves
leaves = sconcat . NonEmpty.map Leaf -- Via Semigroup's sconcat

node :: s -> Tree s a -> Tree s a -> Tree s a -- Combine two trees into an s-labelled node
node = Node

size :: Tree s a -> Int -- The number of leaves in a tree
size = length -- Via Foldable

leafSet :: Ord a => Tree s a -> Set a -- The set of leaves of a tree
leafSet = fold Set.singleton (const Set.union)

hasLeaf :: Eq a => a -> Tree s a -> Bool -- Test if a tree contains a leaf a
hasLeaf = elem -- Via Foldable

prune :: Tree s (Maybe a) -> Maybe (Tree s a) -- Prune all leaves labelled with Nothing
prune = sequence -- Via Traversable

filter :: (a -> Bool) -> Tree s a -> Maybe (Tree s a) -- Prune all leaves a with p a = False
filter p = traverse (\a -> if p a then Just a else Nothing) -- Via Traversable
```

It is sometimes necessary to be able to express the *empty tree*. To do that, we simply wrap the Tree data type into Maybe. This is illustrated by the function prune (Listing 2), which prunes all leaves labelled with Nothing and returns a possibly empty tree as a result. By combining the empty tree and the operation ⋄, we can extend our type of trees from a semigroup to a *monoid*, i.e., a semigroup with a unit element.

When tree interpretations respect the associativity of ⋄, ⊚-labelled trees become isomorphic to non-empty lists. This is evidenced by the functions toList (provided by the Foldable instance in Listing 1) and leaves (Listing 2). In the subsequent sections, we will study data structures obtained by instantiating Tree s a with different semirings s, and by adding more requirements (on top of associativity) to tree interpretations.

## 3   Sets

In this section we instantiate Tree s a with the *trivial semiring* s = () where ⊚ = ① = ():

```haskell
type TSet a = Tree () a -- We prepend T to avoid a name clash with the standard Set
```

As hinted by the type synonym's name, we will interpret values TSet a as (non-empty) sets; to get a standard Set a from a TSet a, we can simply reuse leafSet from Listing 2.

**United Monoids**

■ **Listing 3** Implementing a part of the standard Data.Set API with TSet.

```haskell
type TSet a = Tree () a -- Sets are trees with ()-labelled internal nodes

instance Semiring () where -- The trivial semiring with zero = one = ()
    zero    = ()
    () ⊕ () = ()
    one     = ()
    () ⊗ () = ()

singleton :: a -> TSet a                    insert :: a -> TSet a -> TSet a
singleton = Tree.leaf                       insert = (◇) . singleton -- Via Semigroup

delete :: Eq a => a -> TSet a -> Maybe (TSet a) -- The empty set is represented by Nothing
delete a = Tree.filter (/= a)

member :: Eq a => a -> TSet a -> Bool       union :: TSet a -> TSet a -> TSet a
member = Tree.hasLeaf                       union = (◇) -- Via Semigroup

size :: Ord a => TSet a -> Int
size = Set.size . leafSet -- We can't use Tree.size because it uses a non-idempotent fold

cartesianProduct :: TSet a -> TSet b -> TSet (a, b)
cartesianProduct = liftA2 (,) -- Via Applicative

filter :: (a -> Bool) -> TSet a -> Maybe (TSet a) -- The result can be empty
filter = Tree.filter

map :: (a -> b) -> TSet a -> TSet b
map = fmap -- Via Functor
```

Listing 3 shows how to implement a part of the standard set API by reusing the more general Tree-manipulating functions defined earlier in §2.

In addition to associativity, the set interpretation of trees puts two new requirements on every call site of fold leaf node[7]:

- node zero $a$ $b$ = node zero $b$ $a$, i.e., commutativity of ⊚-labelled nodes;
- node zero $a$ $a$ = $a$, i.e., idempotence[8] of ⊚-labelled nodes.

Therefore we can't use Tree.size in the implementation of TSet.size in Listing 3: Tree.size uses integer addition for folding internal nodes, which is not idempotent. And indeed, Tree.size ("b" ◇ "b") = 2 since the tree has *two leaves*, whereas TSet.size ("b" ◇ "b") = 1 since the set described by the tree has only *one element*.

---

[7] Haskell's type system isn't powerful enough to check such requirements, so we should use fold with care. A dependently typed language, such as Agda [28], could help us here.
[8] An operation • is *idempotent* if $a • a = a$.

12:6

## 4 Simplicial Sets

Describing plain sets with trees isn't new or particularly exciting. However, we hope the reader found it instructive to see what happens if we instantiate Tree s a with the simplest semiring s = (). This section continues by trying the next simplest semiring, i.e., the *Boolean semiring* s = Bool, where ⓪ = False, ① = True, ⊕ = (||) and ⊗ = (&&).

A *simplicial set* [8] is a set of *simplices* of various dimensions along with their incidence relation. A 0-*simplex* is a just a point, or a *vertex*, and will correspond to a leaf in our trees. A 1-*simplex* is a pair of connected vertices, or an *edge*; a 2-*simplex* is a filled-in *triangle*; a 3-*simplex* is a solid *tetrahedron*; and so on. Simplicial sets model systems that cannot be decomposed into pairwise relationships of their components. For example, it can be useful to distinguish three truly concurrent events from three events that are concurrent pairwise but constrained by shared resources [32], e.g., three people eating ice-cream with two spoons. The former corresponds to a 2-simplex, while the latter – to a simplicial set with three 1-simplices (a hollow triangle).

To represent simplices of dimension higher than 0, we will *connect* their vertices using ①-labelled internal nodes. The figure below shows the simplicial sets corresponding to the trees "a", "a" → "b", ("a" → "b") → "c" and ("a" ⋄ "b") → "c", where the operator → is defined as x → y = node one x y (see Listing 4).

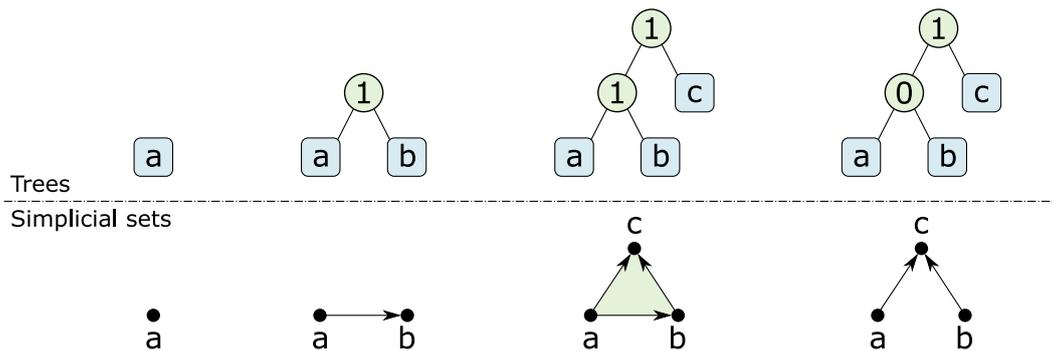

Note the difference between the two last examples: by changing the label of one of the nodes from ① to ⓪, we break the corresponding connection, and as a result, a 2-simplex (triangle *abc*) falls apart into one-dimensional simplices (edges *ac* and *bc*).

When working with simplicial sets, we will extend the set of requirements on interpreting trees by adding the *containment* (see §8) and distributivity laws:

- Associativity, inherited from §2:
  $(a \diamond b) \diamond c = a \diamond (b \diamond c)$  and  $(a \rightarrow b) \rightarrow c = a \rightarrow (b \rightarrow c)$;
- Commutativity and idempotence of ⋄, inherited from §3:
  $a \diamond b = b \diamond a$  and  $a \diamond a = a$;
- Containment (**new!**), i.e., every simplex *contains* its sub-simplices:
  $a \rightarrow b = (a \rightarrow b) \diamond a \diamond b$;
- Distributivity (**new!**):
  $a \rightarrow (b \diamond c) = (a \rightarrow b) \diamond (a \rightarrow c)$  and  $(a \diamond b) \rightarrow c = (a \rightarrow c) \diamond (b \rightarrow c)$.

Adding distributivity is motivated by the desire to make the trees ("a" ⋄ "b") → "c" and ("a" → "c") ⋄ ("b" → "c") mean the same simplicial set (the rightmost one above).

**United Monoids**

◼ **Listing 4**   A basic API for working with simplicial sets represented by trees.

```
type TSimplicialSet a = Tree Bool a -- Simplicial sets are trees over the Boolean semiring

instance Semiring Bool where -- The Boolean semiring with zero = False and one = True
    zero = False
    (⊕)  = (||)
    one  = True
    (⊗)  = (&&)

(→) :: TSimplicialSet a -> TSimplicialSet a -> TSimplicialSet a
(→) = Tree.node one

overlay :: TSimplicialSet a -> TSimplicialSet a -> TSimplicialSet a
overlay = (◇) -- Via Semigroup; recall that (◇) = Tree.node zero

connect :: TSimplicialSet a -> TSimplicialSet a -> TSimplicialSet a
connect = (→)

vertex :: a -> TSimplicialSet a -- 0-simplex
vertex = Tree.leaf

edge :: a -> a -> TSimplicialSet a -- 1-simplex
edge x y = connect (vertex x) (vertex y)

triangle :: a -> a -> a -> TSimplicialSet a      -- 2-simplex
triangle x y z = connect (vertex x) (edge y z) -- Equivalently, connect (edge x y) (vertex z)

simplex :: NonEmpty a -> TSimplicialSet a -- N-simplex
simplex = foldr1 connect . NonEmpty.map vertex -- Via Foldable

hasVertex :: Eq a => a -> TSimplicialSet a -> Bool
hasVertex = Tree.hasLeaf

vertexSet :: Ord a => TSimplicialSet a -> Set a
vertexSet = Tree.leafSet

filter :: (a -> Bool) -> TSimplicialSet a -> Maybe (TSimplicialSet a)
filter = Tree.filter

map :: (a -> b) -> TSimplicialSet a -> TSimplicialSet b -- The so-called "simplicial map"
map = fmap -- Via Functor
```

12:8

Andrey MokhovNow that we know the laws for interpreting trees over the Boolean semiring, we need to choose a target representation for simplicial sets. Alas, there is no standard Data.SimplicialSet, so we need to come up with our own. Here is a simple candidate:

```
type SimplicialSet a = Set (NonEmpty a)
```

Here the inner non-empty lists correspond to simplices, with their vertices listed in the connection order. For example, the simplicial set corresponding to the tree "a" → "b" will contain three lists, one for each contained simplex: [a], [b] and [a,b].

Now we can define vertex, overlay and connect to interpret leaves, ⓪- and ①-labelled nodes, respectively, and use them to fold a Tree Bool a into a SimplicialSet a:

```
vertex :: a -> SimplicialSet a
vertex = Set.singleton . NonEmpty.singleton

overlay :: Ord a => SimplicialSet a -> SimplicialSet a -> SimplicialSet a
overlay = Set.union

connect :: Ord a => SimplicialSet a -> SimplicialSet a -> SimplicialSet a
connect x y =
    Set.unions [x, y, Set.map (uncurry NonEmpty.append) (Set.cartesianProduct x y)]

toSimplicialSet :: Ord a => Tree Bool a -> SimplicialSet a
toSimplicialSet = fold vertex (\s -> if s then connect else overlay)
```

Note that SimplicialSet is a rather naive representation for simplicial sets. In particular, an $n$-simplex contains $O(2^n)$ sub-simplices of lower dimension, and all of them will be included in the outer Set. One can improve SimplicialSet by detecting and storing only maximal simplices or by switching to suffix trees [35]. Thanks to the containment requirement, the Tree-based representation of simplicial sets makes it easy to define an $n$-simplex with a compact $O(n)$-size expression – see the function simplex in Listing 4.

A key advantage of the Tree-based representation is that it is free from any internal invariants, i.e., any value of Tree Bool a describes a (valid) simplicial set. SimplicialSet does not satisfy this property; for example, the set with two lists [a] and [a,b] is not a simplicial set, because the edge *ab* appears in the set without its vertex *b*.

## 5 Graphs

Graphs can be thought of as simplicial sets that contain only 0-simplices (vertices) and 1-simplices (edges). One can therefore represent graphs with a pair of sets:

```
type Graph a = (Set a, Set (a, a)) -- The classic pair (V, E) with invariant E ⊆ V × V
```

To interpret trees Tree Bool a as graphs, we need a way to break up simplices of dimension greater than 1 into edges. To do that, we impose a new folding requirement:

- Decomposition: $a \to b \to c = (a \to b) \diamond (a \to c) \diamond (b \to c)$.

This law is unusual but it is not new: it was introduced as an axiom of *algebraic graphs* in [21]. In fact, by adding this requirement to those used when interpreting simplicial

12:9

**United Monoids**

sets in §4, we get exactly the algebra of non-empty graphs from [21]. It is therefore not surprising that we can reuse more general functions operating on trees and simplicial sets when implementing a part of the API of algebraic graphs – see Listing 5, where hasEdge is the only new function that we needed to implement.

Let us now check that we can fold a Tree Bool a into the corresponding Graph a while respecting decomposition. The last step will be identical to the one from §4:

```
toGraph :: Ord a => Tree Bool a -> Graph a
toGraph = fold vertex (\s -> if s then connect else overlay)
```

All we need to do is adapt vertex, overlay and connect to "truncated" simplicial sets. The truncated variants of vertex and overlay are straightforward:

```
vertex :: a -> Graph a
vertex a = (Set.singleton a, Set.empty)

overlay :: Ord a => Graph a -> Graph a -> Graph a
overlay (v1, e1) (v2, e2) = (Set.union v1 v2, Set.union e1 e2)
```

To connect two graphs, we augment their original edges with a Cartesian product of their vertex sets:

```
connect :: Ord a => Graph a -> Graph a -> Graph a
connect (v1,e1) (v2,e2) = (Set.union v1 v2, Set.unions [e1, e2, Set.cartesianProduct v1 v2])
```

It is not obvious that this interpretation respects the decomposition law, so let us check it. For the vertex set, the law holds because $\cup$ is idempotent:

$$v_1 \cup v_2 \cup v_3 = (v_1 \cup v_2) \cup (v_1 \cup v_3) \cup (v_2 \cup v_3)$$

As for the edge set, one can mechanically check that both sides of the decomposition law evaluate to the following expression when $(\diamond) =$ overlay and $(\rightarrow) =$ connect.

$$e_1 \cup e_2 \cup e_3 \cup (v_1 \times v_2) \cup (v_1 \times v_3) \cup (v_2 \times v_3)$$

### 5.1 Other Kinds of Graphs

So far we discussed only directed graphs with no edge labels. One way to support other kinds of graphs is to strengthen the tree folding requirements. Specifically:

- For *undirected graphs*, $\rightarrow$ must commute for leaves: leaf $a \rightarrow$ leaf $b =$ leaf $b \rightarrow$ leaf $a$. Taken with other requirements, this leads to general commutativity $a \rightarrow b = b \rightarrow a$.
- For *reflexive graphs*, $\rightarrow$ must have no effect on leaves: leaf $a =$ leaf $a \rightarrow$ leaf $a$.
- For *bipartite graphs*, $\rightarrow$ must have no effect on leaves corresponding to vertices from the same part. That is, $\forall (a, b) \in (P \times P) \cup (Q \times Q)$, leaf $a \diamond$ leaf $b =$ leaf $a \rightarrow$ leaf $b$, where $P$ and $Q$ are the two parts of the bipartite graph.
- For *transitively-closed* and *acyclic graphs*, see §6.

Another approach to supporting new kinds of graphs is instantiating Tree s a with other semirings s. In §7, this will allow us to support *edge-labelled graphs*.

Andrey Mokhov■ **Listing 5** Implementing a part of the Algebra.Graph API [21] with Tree Bool a.

```haskell
type TGraph a = Tree Bool a -- Graphs are trees over the Boolean semiring

vertex :: a -> TGraph a
vertex = SimplicialSet.vertex

edge :: a -> a -> TGraph a
edge = SimplicialSet.edge

vertices :: NonEmpty a -> TGraph a
vertices = Tree.leaves

clique :: NonEmpty a -> TGraph a -- A clique is a set of pairwise connected vertices
clique = SimplicialSet.simplex

overlay :: TGraph a -> TGraph a -> TGraph a
overlay = SimplicialSet.overlay

connect :: TGraph a -> TGraph a -> TGraph a
connect = SimplicialSet.connect

vertexSet :: Ord a => TGraph a -> Set a
vertexSet = SimplicialSet.vertexSet

edgeSet :: Ord a => TGraph a -> Set (a, a)
edgeSet = snd . toGraph

induce :: (a -> Bool) -> TGraph a -> Maybe (TGraph a) -- The result may be empty
induce = Tree.filter

hasVertex :: Eq a => a -> TGraph a -> Bool
hasVertex = SimplicialSet.hasVertex

hasEdge :: Ord a => a -> a -> TGraph a -> Bool
hasEdge x y graph = case induce (\a -> a == x || a == y) graph of
    Nothing  -> False
    Just sub -> elem (x, y) (edgeSet sub) -- Here sub can only contain vertices x and y
```

12:11

## 6 Preorders and Generalised Transitively-Closed Graphs

A *preorder* is a binary relation $\preceq$ that is:

- *Reflexive*: $a \preceq a$; and
- *Transitive*: $a \preceq b \wedge b \preceq c \Rightarrow a \preceq c$.

As mentioned earlier in §5.1, reflexivity can be expressed as a new law, which requires that connecting a leaf to itself must be redundant:

- Reflexivity: leaf $a$ = leaf $a \rightarrow$ leaf $a$.

To ensure that preorder expressions are interpreted "modulo transitive connections", we also add the following transitivity law. It states that adding/removing a transitive connection $a \rightarrow c$ to/from the tree $a \rightarrow b \diamond b \rightarrow c$ must be redundant:

- Transitivity: $a \rightarrow b \diamond b \rightarrow c = a \rightarrow b \diamond b \rightarrow c \diamond a \rightarrow c$.

Reading the reflexivity and transitivity laws from left to right, one can think of them as introducing the preorder terms $a \preceq a$ and $a \preceq c$ under suitable assumptions.

We can generalise this approach to trees of type Tree s a where s is an arbitrary semiring. For clarity, let us adopt the following shorthand notation:

$$a \xrightarrow{x} b = \text{node } x \text{ } a \text{ } b$$

Then generalised reflexivity and transitivity laws can be expressed as follows.

- *Unit*: leaf $a$ = leaf $a \xrightarrow{①}$ leaf $a$;
- *Composition*: $a \xrightarrow{x} b \diamond b \xrightarrow{y} c = a \xrightarrow{x} b \diamond b \xrightarrow{y} c \diamond a \xrightarrow{x \otimes y} c$.

Recalling that semiring's operator $\otimes$ composes connectivities in sequence and that it has ① as the unit, we can refer to these two laws together as *sequential composition*.

These laws might remind you of the initialisation and relaxation steps used in various shortest-path finding algorithms. More precisely, in the *tropical semiring* [20], where $⓪ = \infty$, $① = 0$, $\oplus = \min$, and $\otimes = +$, the unit law corresponds to initialising the distance from a vertex to itself to 0, and composition to relaxing the distance $a \rightarrow c$ to $x + y$ after finding paths $a \rightarrow b$ and $b \rightarrow c$ with distances $x$ and $y$, respectively.

Section §7 will introduce a few missing tools for working with generalised edge-labelled graphs. With these tools, and the Floyd-Warshall-Kleene algorithm [15][18], we will be able to interpret trees over an arbitrary semiring and satisfy the sequential composition requirements. See an implementation in [23].

### 6.1 Acyclic Graphs and Partial Orders

The Tree data type cannot rule out cyclic terms like $a \rightarrow b \diamond b \rightarrow a$ at the type level. If these terms are undesirable, we can turn them all to *errors* by adding two new laws:

- *Cycle*: leaf $a \rightarrow$ leaf $a = \bot$, where $\bot$ is a "cycle error" value in the target type.
- *Zero*: $a \diamond \bot = a \rightarrow \bot = \bot \rightarrow a = \bot$.

These laws plus transitivity ensure that all cyclic terms are equivalent to $\bot$. This allows us to represent strict partial orders that assert $\neg(a \prec a)$, and more general concurrency models such as partial order relations labelled with Boolean predicates [22].

## 7 Edge-Labelled Graphs

Equipped with the notation from the previous section §6, we can use our tree language to describe graphs whose edges are labelled with values from an arbitrary semiring. For example, the expression "a" $\xrightarrow{①}$ "b" $\diamond$ "b" $\xrightarrow{②}$ "c" corresponds to the graph below.

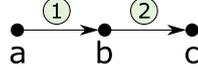

To give meaning to parallel edges, we introduce the *parallel composition* law. It is similar to the sequential composition from §6 but concerns ⊚ and ⊕ instead of ① and ⊗:

- *Unit*: leaf $a$ = leaf $a \xrightarrow{⊚}$ leaf $a$;
- *Composition*: $a \xrightarrow{x} b \diamond a \xrightarrow{y} b = a \xrightarrow{x \oplus y} b$.

Note that if the semiring is idempotent, i.e., $x \oplus x = x$, then these laws imply general idempotence $a \diamond a = a$. Indeed, if $a$ is a leaf, then idempotence is just the unit law. Otherwise, if $a = p \xrightarrow{x} q$, then idempotence follows from the composition law:

$$p \xrightarrow{x} q \diamond p \xrightarrow{x} q = p \xrightarrow{x \oplus x} q = p \xrightarrow{x} q.$$

Furthermore, containment (§4) is also a special case of parallel composition:

$$\begin{aligned}
a \to b \diamond a \diamond b &= a \xrightarrow{①} b \diamond a \xrightarrow{⊚} b & \text{(definitions of } \to, \diamond \text{ and } \xrightarrow{x}) \\
&= a \xrightarrow{① \oplus ⊚} b & \text{(parallel composition)} \\
&= a \xrightarrow{①} b & \text{(unit of } \oplus) \\
&= a \to b & \text{(definitions of } \to \text{ and } \xrightarrow{x})
\end{aligned}$$

Another tool that we need for interpreting edge-labelled graphs is the generalised decomposition law, which gives meaning to nested nodes.

- *Decomposition*: $\begin{cases} (a \xrightarrow{x} b) \xrightarrow{y} c = a \xrightarrow{x} b \diamond a \xrightarrow{y} c \diamond b \xrightarrow{y} c; \\ a \xrightarrow{x} (b \xrightarrow{y} c) = a \xrightarrow{x} b \diamond a \xrightarrow{x} c \diamond b \xrightarrow{y} c. \end{cases}$

Interestingly, decomposition subsumes both associativity and distributivity. Indeed, to derive associativity, we can instantiate both equalities with $x = y$:

$$(a \xrightarrow{x} b) \xrightarrow{x} c = a \xrightarrow{x} b \diamond a \xrightarrow{x} c \diamond b \xrightarrow{x} c = a \xrightarrow{x} (b \xrightarrow{x} c).$$

Distributivity then follows from decomposition by setting the inner label to ⊚.

$$\begin{aligned}
(a \diamond b) \xrightarrow{x} c &= (a \xrightarrow{⊚} b) \xrightarrow{x} c & \text{(definitions of } \diamond \text{ and } \xrightarrow{x}) \\
&= a \xrightarrow{⊚} b \diamond a \xrightarrow{x} c \diamond b \xrightarrow{x} c & \text{(left decomposition)} \\
&= a \diamond b \diamond a \xrightarrow{x} c \diamond b \xrightarrow{x} c & \text{(definitions of } \diamond \text{ and } \xrightarrow{x}) \\
&= (a \diamond a \xrightarrow{x} c) \diamond (b \diamond b \xrightarrow{x} c) & \text{(associativity and commutativity of } \diamond) \\
&= a \xrightarrow{x} c \diamond b \xrightarrow{x} c & \text{(containment)}
\end{aligned}$$

To conclude, the complete set of requirements for interpreting edge-labelled graphs is: commutativity of $\diamond$ inherited from sets (§3), parallel composition and decomposition.

**United Monoids**

With all the folding requirements in place, we can interpret Tree expressions into the following "flat" representation of edge-labelled graphs, which maps vertices a to a collection of s-labelled outgoing edges.

```
type LGraph s a = Map a (Map a s)
```

Creating singleton graphs and overlaying graphs is fairly straightforward:

```
vertex :: a -> LGraph s a
vertex x = Map.singleton x Map.empty

overlay :: (Ord a, Semiring s) => LGraph s a -> LGraph s a -> LGraph s a
overlay = Map.unionWith (Map.unionWith (⊕)) -- Parallel composition
```

When implementing connect, we come across a subtle issue with LGraph: it is possible to unintentionally accumulate redundant ⊙-labelled edges in the inner Map. Filtering out these redundant edges from LGraph s a is an important optimisation when working with sparse graphs in practice, which is why the code below requires Eq s.[9] Note that the Tree-based representation of graphs sidesteps this issue because overlay x y and connect zero x y are represented by the same tree Node zero x y.

```
connect :: (Ord a, Semiring s, Eq s) => s -> LGraph s a -> LGraph s a -> LGraph s a
connect s x y
    | s == zero = overlay x y -- Filter out redundant zero-labelled edges
    | otherwise = Map.unionsWith (Map.unionWith (⊕)) [ x, y, newEdges ]
  where
    outgoing = Map.fromSet (const s) (Map.keysSet y)
    newEdges = Map.fromSet (const outgoing) (Map.keysSet x) -- Decomposition
```

Finally, we can fold a Tree s a to turn it into the corresponding LGraph s a:

```
toLGraph :: (Ord a, Semiring s, Eq s) => Tree s a -> LGraph s a
toLGraph = fold vertex connect
```

It is useful to combine edge-labelled graphs with the sequential composition law described in the previous section (§6). By computing the reflexive and transitive closure of an edge-labelled graph, it is possible to solve a variety of semiring optimisation problems, e.g., see [6][20]. An implementation is available in [23].

We have now seen how to use trees as a language for describing and manipulating various kinds of sets and graphs. The next section introduces *united monoids* that provide a common ground for these seemingly different combinatorial data structures.

---

[9] We do not need the full power of Eq s here: we just need to test if a given edge label is ⊙. For some types s, testing for ⊙ is much easier than a general equality test; for example, testing if a Map k v is empty is trivial but Eq (Map k v) is undecidable for some choices of v. To resolve this problem, one can switch to a more precise constraint like EqZero s.

## 8 United Monoids

In this section we introduce *united monoids* as an algebraic structure that turns out to be a common ground for the graph-like structures discussed earlier in §3-§7.

A *monoid* $(S, \diamond, \varepsilon)$ is a way to express a basic form of composition in mathematics: any two elements $a$ and $b$ of the set $S$ can be composed into a new element $a \diamond b$ of the same set $S$, and, furthermore, there is a special element $\varepsilon \in S$, which is the unit element of the composition, as expressed by the *unit axioms* $a \diamond \varepsilon = \varepsilon \diamond a = a$. In words, composing the unit element with another element does not change the latter.

Monoids often come in pairs: addition and multiplication $(+, \times)$, disjunction and conjunction $(\vee, \wedge)$, set union and intersection $(\cup, \cap)$, parallel and sequential composition of processes $(|, ;)$, etc. Two common ways in which such monoid pairs can form are called *semirings* and *lattices*. In fact, the former have played a major role in this paper so far. Below we briefly introduce the latter.

A *bounded lattice* $(S, \vee, 0, \wedge, 1)$ comprises two monoids, which are called *join* $(S, \vee, 0)$ and *meet* $(S, \wedge, 1)$. They operate on the same set, are required to be commutative and idempotent, and satisfy the following *absorption axioms*: $a \wedge (a \vee b) = a \vee (a \wedge b) = a$. Like semirings, lattices show up very frequently in different application areas.

### 8.1 What if $0 = 1$?

What happens when the units of the two monoids in a pair coincide, i.e., when $0 = 1$? In a semiring $(S, \oplus, ⓞ, \otimes, ①)$, this leads to devastating consequences. Not only ① becomes equal to ⓞ, but all other elements of the semiring become equal to ⓞ too, as demonstrated below.

$$
\begin{aligned}
a &= ① \otimes a && \text{(unit of } \otimes\text{)} \\
&= ⓞ \otimes a && \text{(we postulate } ⓞ = ①\text{)} \\
&= ⓞ && \text{(zero of } \otimes\text{)}
\end{aligned}
$$

That is, the semiring is annihilated into a single point ⓞ, becoming isomorphic to the trivial semiring `s = ()`, which we have come across in §3.

In a bounded lattice $(S, \vee, 0, \wedge, 1)$, postulating $0 = 1$ leads to the same catastrophe, albeit in a different manner:

$$
\begin{aligned}
a &= 1 \wedge a && \text{(unit of } \wedge\text{)} \\
&= 1 \wedge (0 \vee a) && \text{(unit of } \vee\text{)} \\
&= 0 \wedge (0 \vee a) && \text{(we postulate } 0 = 1\text{)} \\
&= 0 && \text{(absorption)}
\end{aligned}
$$

That is, the lattice is absorbed into a single point 0.

Postulating $0 = 1$ has so far led to nothing but disappointment. In the next subsection we find another way of pairing monoids, which does not involve the axioms of annihilation and absorption, and makes the resulting structure more interesting.

**United Monoids**

### 8.2 From $0 = 1$ to Containment Laws and Back

Consider two monoids $(S, +, 0)$ and $(S, \cdot, 1)$, such that $+$ is commutative and $\cdot$ distributes over $+$. We call these monoids *united* if $0 = 1$. To avoid confusion with semirings and lattices, we will use $\varepsilon$ to denote the unit element of both monoids, that is, $a + \varepsilon = a \cdot \varepsilon = a$. Note: we will often omit the operator $\cdot$ and write simply $ab$ instead of $a \cdot b$, which is a common convention. We will further refer to $\varepsilon$ as *empty*, the operation $+$ as *overlay*, and the operation $\cdot$ as *connect*.

What can we tell about united monoids? First of all, it is easy to show that the monoid $(S, +, \varepsilon)$ is idempotent:

$$
\begin{aligned}
a + a &= a\varepsilon + a\varepsilon && \text{(unit of }\cdot\text{)} \\
&= a(\varepsilon + \varepsilon) && \text{(distributivity)} \\
&= a\varepsilon && \text{(unit of }+\text{)} \\
&= a && \text{(unit of }\cdot\text{)}
\end{aligned}
$$

This means $(S, +, \varepsilon)$ is a commutative idempotent monoid, i.e., a *bounded semilattice*.

The next property of is more unusual: $ab = ab + a = ab + b = ab + a + b$. We call these equalities the *containment laws*: intuitively, when you connect $a$ and $b$, the constituent parts are contained in the result $ab$. Let us prove containment:

$$
\begin{aligned}
ab + a &= ab + a\varepsilon && \text{(unit of }\cdot\text{)} \\
&= a(b + \varepsilon) && \text{(distributivity)} \\
&= ab && \text{(unit of }+\text{)}
\end{aligned}
$$

The two other laws are proved analogously (in fact, they are equivalent to each other).

Surprisingly, the containment law $ab = ab + a$ is equivalent to $0 = 1$, i.e., the latter can also be proved from the former:

$$
\begin{aligned}
0 &= 1 \cdot 0 && \text{(1 is the unit of }\cdot\text{)} \\
&= 1 \cdot 0 + 1 && \text{(containment)} \\
&= 0 + 1 && \text{(1 is the unit of }\cdot\text{)} \\
&= 1 && \text{(0 is the unit of }+\text{)}
\end{aligned}
$$

This means that united monoids can be equivalently characterised by the containment laws, which makes it possible to also talk about *united semigroups*. The term "united semigroup" may seem somewhat nonsensical (semigroups have no "units" after all), however, note that if they secretly had unit elements, they would have to coincide.

In the same manner, the containment laws imply that zeroes of the two operations must be the same. Let $z^+$ and $z^{\cdot}$ denote the zeroes of $+$ and $\cdot$, respectively. Then:

$$
\begin{aligned}
z^+ &= z^{\cdot} + z^+ && (z^+ \text{ is the zero of } +) \\
&= z^{\cdot} \cdot z^+ + z^+ && (z^{\cdot} \text{ is the zero of } \cdot) \\
&= z^{\cdot} \cdot z^+ && \text{(containment)} \\
&= z^{\cdot} && (z^{\cdot} \text{ is the zero of } \cdot)
\end{aligned}
$$

Back in 6.1, we used zero $\bot$ to model "cycle errors". As we now see, requiring both $a \diamond \bot = a$ and $a \to \bot = \bot \to a = \bot$ was unnecessary: we could have picked just one of these laws and the other one would follow. For general (i.e., not acyclic) graphs, the fully-connected graph acts as the zero.

Finally, let us prove another unusual property of united monoids: overlay and connect can have no inverses with respect to $\varepsilon$. That is: if $a + b = \varepsilon$ or $ab = \varepsilon$ then $a = b = \varepsilon$.

$$
\begin{array}{rll}
a &=& a + \varepsilon \quad \text{(unit of +)} \\
  &=& a + a + b \quad \text{(assumption } a + b = \varepsilon\text{)} \\
  &=& a + b \quad \text{(idempotence of +)} \\
  &=& \varepsilon \quad \text{(assumption } a + b = \varepsilon\text{)}
\end{array}
\qquad
\begin{array}{rll}
a &=& a + \varepsilon \quad \text{(unit of +)} \\
  &=& a + ab \quad \text{(assumption } ab = \varepsilon\text{)} \\
  &=& ab \quad \text{(containment)} \\
  &=& \varepsilon \quad \text{(assumption } ab = \varepsilon\text{)}
\end{array}
$$

Therefore, it is not possible to extend united monoids to ring- or field-like structures with inverses.

It is time to look at some examples of united monoids.

### 8.3 Examples from This Paper

The containment laws should have reminded you of simplicial sets since the latter are closed in terms of containment. For example, a filled-in triangle contains its edges and vertices, and it cannot appear in a simplicial set without any of them. This property can be expressed algebraically as: $abc = abc + ab + ac + bc + a + b + c$. Interestingly, this "3D" containment law follows from the "2D" version for united monoids:

$$
\begin{array}{rll}
abc &=& (ab + a + b)c \quad \text{(containment)} \\
    &=& (ab)c + ac + bc \quad \text{(distributivity)} \\
    &=& (abc + ab + c) + (ac + a) + (bc + b) \quad \text{(containment)} \\
    &=& abc + ab + ac + bc + a + b + c \quad \text{(commutativity)}
\end{array}
$$

We can similarly prove higher-dimensional versions of the containment law; they all follow from the basic law $ab = ab + a$, or, alternatively, from $0 = 1$.

Unlabelled graphs, preorders, and partial orders also satisfy the containment laws. In this paper, we focused on non-empty structures, which is why we didn't come across units of the overlay and connect operators. By wrapping trees in Maybe, as we've done in a few cases, we were essentially augmenting our structures with the unit Nothing.

It is worth remarking on the two remaining cases: sets and edge-labelled graphs. When working with sets, we had only one operator $\diamond$, which was a consequence of the fact that we used a trivial semiring with $\circledcirc = \circledone$. One can think of this case as the monoid $\diamond$ being united with itself.

The case of edge-labelled graphs is more interesting. There we had as many operations as there were elements in the edge label semiring s. It turns out that every monoid $\xrightarrow{x}$ is united with the overlay monoid $\diamond$. In other words, edge-labelled graphs form a *semiring of united monoids*.

### 8.4 Other Examples

United monoids also appear when studying the composition of concurrent processes. As an example, consider Haskell's language extension ApplicativeDo [19], which uses a simple cost model for characterising the execution time of programs that are composed in parallel and in sequence.

**United Monoids**

The cost model comprises two monoids:

- $(\mathbb{Z}^{\geq 0}, \max, 0)$: the execution time of programs $a$ and $b$ composed in parallel is defined to be the maximum of their execution times:

    $\text{time}(a \mid b) = \max(\text{time}(a), \text{time}(b))$

- $(\mathbb{Z}^{\geq 0}, +, 0)$: the execution time of programs $a$ and $b$ composed in sequence is defined to be the sum of their execution times:

    $\text{time}(a \,;\, b) = \text{time}(a) + \text{time}(b)$

Execution times are non-negative, hence both max and + have unit 0, which is the execution time of the *empty program*. It is easy to check that distributivity (+ distributes over max) and containment laws hold. Note that the resulting algebraic structure is different from the tropical max-plus semiring $(\mathbb{R}^{-\infty}, \max, -\infty, +, 0)$ commonly used in scheduling, where the unit of max is $-\infty$ but the unit of + is 0.

Interestingly, parallel and sequential composition of programs also forms a united monoid, where the empty program is the unit. One can therefore call the cost model function time a *united monoid homomorphism*.[10] Another related example can be found in [1], where asynchronous circuit specifications are composed in parallel and in sequence, using graphs labelled with the semiring of Boolean predicates.

## 9 Related Work

This paper continues the research on *algebraic graphs* started in [21]. Since their introduction, algebraic graphs have been implemented in several languages, including Agda [25], F# [5], Haskell [23], Kotlin [29], PureScript [14], Scala [26], TypeScript [2], and R [27]. Algebraic graphs also found applications in industry, e.g., in GitHub's static program analysis library Semantic [3][11]. In this paper we further distil the essence of the algebraic approach to working with graphs by replacing the data type

```
data Graph a = Empty | Vertex a | Overlay (Graph a) (Graph a) | Connect (Graph a) (Graph a)
```

with binary trees parameterised by a semiring. We remove the Empty constructor to make it possible to distinguish non-empty structures at the type level and also to keep expressions free from redundant empty leaves. By merging the Overlay and Connect constructors into a single Node parameterised by a semiring, we get a more general data structure. This allows us to resolve one of the main limitations of the work in [21], which was the lack of support for edge labels.

Two other popular approaches for representing graphs in functional programming languages were developed by King and Launchbury [17] (based on adjacency lists, available from the containers library) and Erwig [7] (based on *inductive graphs*, where graphs can be decomposed into a *context*, i.e., a vertex with its neighbourhood, and the rest of the graph). These and other classic graph representations can be obtained from our tree expressions via folding. Note that, as we remarked in §4, folding a tree expression into a "flat" graph representation too early can be undesirable because of the loss of compactness and the need to maintain various correctness invariants.

---

[10] A *homomorphism* is a structure-preserving map between two algebraic structures.

Using semirings as a general framework for solving problems on graphs is, of course, not new; see, for instance, [20]. In the functional programming community, Dolan's "Fun with Semirings" [6] is a notable example of parameterising matrices with semirings, which motivated the author of this paper to apply semirings in the context of algebraic graphs. In a more recent work, Kidney and Wu [16] use semirings for generalising algorithms for weighted search on graphs represented by functions from vertices to their neighbours.

Jeremy Gibbons defines operations above and beside for composing directed acyclic graphs [9]. They have the same unit (the empty graph) but the operation beside is defined only for graphs of matching types. One can therefore think of above and beside as *partial united monoids* or *united categories*. Similarly, Santos and Oliveira [33] show how to compose type-compatible matrices vertically and horizontally using the operations fork and join, respectively, whose common unit is the empty matrix.

Various flavours of parallel and sequential composition often form united monoids. In the paper on Concurrent Kleene Algebra [13], Hoare et al. use the term "bimonoid" to refer to such structures (without investigating them in detail). In category theory, "bimonoid" is used for an unrelated concept (a structure with a monoid and a comonoid [30]). The author of this paper chose to use the term "united monoids", because it appears to be unused and is also more specific compared to "bimonoid".

Rivas and Jaskelioff discuss various notions of computation viewing them as monoids in the category of endofunctors [31]. Two of these monoids, namely Applicative and Monad, turn out to have the same unit pure = return, and they can therefore be considered *united monoids in the category of endofunctors*. Investigating categorical equivalents of the presented ideas is an interesting direction of future work.

**Acknowledgements**   I would like to thank everyone who contributed to this research by giving feedback on earlier versions of this work (particularly, [21] and [24]), participating in numerous online and in-person discussions, and last but not least, expressing genuine interest and encouragement. Here is a most likely incomplete list (I apologise for any omissions): Arseniy Alekseyev, Roland Backhouse, Gershom Bazerman, Dave Clarke, Ulan Degenbaev, Dominique Devriese, Chris Doran, Victor Khomenko, Adithya Kumar, Dave Long, Anton Lorenzen, Georgy Lukyanov, Alp Mestanogullari, Alexandre Moine, Andreas Nuyts, Greg Pfeil, Stefan Plantikow, Armando Santos, Danil Sokolov, Sjoerd Visscher, Leo White and Alex Yakovlev. I am also grateful to the Programming Journal's Programme Committee for the thorough reviews that helped me improve the final version of the paper.

**United Monoids**

**References**

[1] Jonathan Beaumont, Andrey Mokhov, Danil Sokolov, and Alex Yakovlev. "High-level asynchronous concepts at the interface between analog and digital worlds". In: *IEEE Transactions on Computer-Aided Design of Integrated Circuits and Systems* 37.1 (2017). https://doi.org/10.1109/TCAD.2017.2748002, pages 61–74.

[2] Yuriy Bogomolov. "Implementation of algebraic graphs in TypeScript". https://github.com/algebraic-graphs/typescript. 2020, last accessed on 12 January 2022.

[3] Timothy Clem and Patrick Thomson. "Static Analysis at GitHub". In: *ACM Queue* 19.4 (2021). https://doi.org/10.1145/3487019.3487022.

[4] Thomas H. Cormen, Charles E. Leiserson, Ronald L. Rivest, and Clifford Stein. *Introduction to Algorithms*. MIT Press, 2009. ISBN: 9780262258104.

[5] Nicholas Cowle. "Implementation of algebraic graphs in F#". https://github.com/algebraic-graphs/fsharp. 2017, last accessed on 12 January 2022.

[6] Stephen Dolan. "Fun with semirings: a functional pearl on the abuse of linear algebra". In: *ACM SIGPLAN Notices*. Volume 48. 9. https://doi.org/10.1145/2544174.2500613. ACM. 2013, pages 101–110.

[7] Martin Erwig. "Inductive graphs and functional graph algorithms". In: *Journal of Functional Programming* 11.05 (2001). https://doi.org/10.1017/S0956796801004075, pages 467–492.

[8] Greg Friedman. "Survey Article: An elementary illustrated introduction to simplicial sets". In: *Rocky Mountain Journal of Mathematics* 42.2 (2012). https://doi.org/10.1216/RMJ-2012-42-2-353, pages 353–423.

[9] Jeremy Gibbons. "An initial-algebra approach to directed acyclic graphs". In: *International Conference on Mathematics of Program Construction*. https://dl.acm.org/doi/10.5555/648083.747153. Springer. 1995, pages 282–303.

[10] Jeremy Gibbons and Nicolas Wu. "Folding Domain-Specific Languages: Deep and Shallow Embeddings (Functional Pearl)". In: *Proceedings of the 19th ACM SIGPLAN International Conference on Functional Programming*. https://doi.org/10.1145/2628136.2628138. ACM, 2014, pages 339–347.

[11] GitHub. "Semantic project". https://github.com/github/semantic. 2019, last accessed on 12 January 2022.

[12] Jonathan S. Golan. *Semirings and their Applications*. Springer Science & Business Media, 1999.

[13] Tony Hoare, Bernhard Möller, Georg Struth, and Ian Wehrman. "Concurrent Kleene algebra and its foundations". In: *The Journal of Logic and Algebraic Programming* 80.6 (2011). https://doi.org/10.1016/j.jlap.2011.04.005, pages 266–296.

[14] Thomas Honeyman. "Implementation of algebraic graphs in PureScript". https://github.com/thomashoneyman/purescript-alga. 2019, last accessed on 12 January 2022.

[15]  John E. Hopcroft and Jeffrey D. Ullman. *Introduction to Automata Theory, Languages, and Computation*. Addison-Wesley, 1979.

[16]  Donnacha Oisin Kidney and Nicolas Wu. "Algebras for Weighted Search". In: *Proc. ACM Program. Lang.* 5.ICFP (2021). https://doi.org/10.1145/3473577.

[17]  David J. King and John Launchbury. "Structuring depth-first search algorithms in Haskell". In: *Proceedings of the 22nd ACM SIGPLAN-SIGACT symposium on Principles of programming languages*. https://doi.org/10.1145/199448.199530. ACM. 1995, pages 344–354.

[18]  Stephen Cole Kleene. *Representation of Events in Nerve Nets and Finite Automata*. Technical report. RAND PROJECT AIR FORCE SANTA MONICA CA, 1951.

[19]  Simon Marlow, Simon Peyton Jones, Edward Kmett, and Andrey Mokhov. "Desugaring Haskell's Do-Notation into Applicative Operations". In: *Proceedings of the 9th International Symposium on Haskell*. https://doi.org/10.1145/2976002.2976007. ACM, 2016, pages 92–104.

[20]  Mehryar Mohri. "Semiring frameworks and algorithms for shortest-distance problems". In: *Journal of Automata, Languages and Combinatorics* 7.3 (2002). https://dl.acm.org/doi/10.5555/639508.639512, pages 321–350.

[21]  Andrey Mokhov. "Algebraic Graphs with Class (Functional Pearl)". In: *Proceedings of the 10th ACM SIGPLAN International Symposium on Haskell*. https://doi.org/10.1145/3122955.3122956. ACM, 2017, pages 2–13.

[22]  Andrey Mokhov. "Conditional partial order graphs". PhD thesis. Newcastle University, 2009.

[23]  Andrey Mokhov. "Implementation of algebraic graphs in Haskell". https://github.com/snowleopard/alga. 2016, last accessed on 12 January 2022.

[24]  Andrey Mokhov. "United Monoids". Early results published in a blog post https://blogs.ncl.ac.uk/andreymokhov/united-monoids/ along with an implementation and proofs https://github.com/snowleopard/united. 2018, last accessed on 12 January 2022.

[25]  Andrey Mokhov, Arseniy Alekseyev, Anton Lorenzen, and Alexandre Moine. "Implementation of algebraic graphs in Agda". https://github.com/algebraic-graphs/agda. 2017, last accessed on 12 January 2022.

[26]  Andrey Mokhov and Ivan Poliakov. "Implementation of algebraic graphs in Scala". https://github.com/algebraic-graphs/scala. 2017, last accessed on 12 January 2022.

[27]  Ian Moran. "Implementation of algebraic graphs in R". https://github.com/algebraic-graphs/R. 2021, last accessed on 12 January 2022.

[28]  Ulf Norell. "Towards a practical programming language based on dependent type theory". PhD thesis. Chalmers University of Technology, 2007.

[29]  Alexandre Piveteau. "Implementation of algebraic graphs in Kotlin". https://github.com/alexandrepiveteau/algebraic-graphs-kotlin. 2018, last accessed on 12 January 2022.

**United Monoids**

[30] Hans-E. Porst. "On Categories of Monoids, Comonoids, and Bimonoids". In: *Quaestiones Mathematicae* 31.2 (2008), pages 127–139.

[31] Exequiel Rivas and Mauro Jaskelioff. "Notions of computation as monoids". In: *Journal of Functional Programming* 27 (2017). https://doi.org/10.1017/S0956796817000132, e21.

[32] Leonid Rosenblum, Alexandre Yakovlev, and Vladimir Yakovlev. "A look at concurrency semantics through "lattice glasses"". In: *Bulletin of the EATCS* 37 (1989), pages 175–180.

[33] Armando Santos and José N. Oliveira. "Type Your Matrices for Great Good: A Haskell Library of Typed Matrices and Applications (Functional Pearl)". In: *Proceedings of the 13th ACM SIGPLAN International Symposium on Haskell*. https://doi.org/10.1145/3406088.3409019. ACM, 2020, pages 54–66.

[34] Peter Selinger. "A survey of graphical languages for monoidal categories". In: *New structures for physics*. Springer, 2010, pages 289–355.

[35] Peter Weiner. "Linear pattern matching algorithms". In: *14th Annual Symposium on Switching and Automata Theory*. https://doi.org/10.1109/SWAT.1973.13. 1973, pages 1–11.

## About the author

**Andrey Mokhov** is a software engineer at Jane Street London, and a visiting fellow at Newcastle University, UK. His research interests are in applying abstract mathematics and functional programming to solving large-scale engineering problems.
**email:** andrey.mokhov@ncl.ac.uk